\documentclass[twocolumn,aps,showpacs]{revtex4}
\usepackage{graphicx}

\newcommand{\tht}{\vartheta}
\def\be{\begin{equation}}
\def\ee{\end{equation}}
\def\ba{\begin{eqnarray}}
\def\ea{\end{eqnarray}}
\def\de{\partial}
\def\div{\nabla\cdot}
\def\grad{\nabla}

\def\ltsima{$\; \buildrel < \over \sim \;$}
\def\simlt{\lower.5ex\hbox{\ltsima}}
\def\gtsima{$\; \buildrel > \over \sim \;$}
\def\simgt{\lower.5ex\hbox{\gtsima}}
\begin{document}
\title{\bf Radiation-condensation instability in a four-fluid dusty plasma}
\author{Andreas Kopp\thanks{ak@tp4.ruhr-uni-bochum.de}}
\affiliation{Institut f\"ur Theoretische Physik IV and Astronomisches
Institut, Ruhr-Universit\"at Bochum, D-44780 Bochum, Germany}
\author{Yuri A. Shchekinov\thanks{yus@phys.rsu.ru}}
\affiliation{Department of Physics, South Federal University, Rostov on Don,
344090 Russia}

\begin{abstract}
In this work linear stability analysis of a four-fluid optically thin plasma
consisting of electrons, ions, neutral atoms, and charged dust particles is
performed with respect to the radiation-condensation (RC) instability. The
energy budget of the plasma involves the input from  heating through
photo-electron emission by dust particles under external ultraviolet radiation
as well as radiative losses in inelastic electron-neutral, electron-ion,
neutral-neutral collisions. It is shown that negatively charged particles
stimulate the RC instability in the sense that the conditions for the
instability to hold are wider than similar conditions in a single-fluid
description. 
\end{abstract}
\maketitle

\section{INTRODUCTION}

The radiation-condensation (RC) instability in an optically thin plasma plays
an important role both in space and laboratory plasmas
(cf. Refs.~\onlinecite{field,meerson,brk,psetal}). Dust impurities can alter
dynamical and thermodynamical properties of the plasma because of their high
inertia and the ability to transform  thermal energy into radiation. It was
shown by Ref. \onlinecite{ibanez} within the framework of a two-fluid (dust
and plasma) description that the presence of dust particles can change the
conditions for the RC instability to grow. An essential assumption of
Ref.~\onlinecite{ibanez} was that dust particles contribute to the radiative
cooling. In many situations, however, dust provides instead the energy input
to plasma due to photo-emission of electrons from the dust surface when the
system is exposed to ultraviolet radiation. Such conditions can be met in
interstellar plasmas (cf. Ref.~\onlinecite{tielens}). In this case the energy
gained from non-thermal electrons escaping photo-ionized dust grains is
transmitted to thermal electrons in the ambient plasma, which then in turn heat
ions and neutrals in elastic collisions. Simultaneously, the electrons lose
their energy and pressure radiatively, and therefore are compressed under the
external pressure. In this picture dust particles and electrons are the
thermodynamically active components, while neutrals and ions represent the
passive components providing conditions for radiative cooling of the
electrons. Therefore, at such circumstances only a full four-fluid description
is adequate. For the sake of simplicity we restrict ourself in this paper to 
consideration of a non-magnetized plasma. 

The paper is organized in the following way: in Sec.~II.A we present the
full set of four-fluid equations describing the evolution of the 
plasma under consideration, followed by the description of the 
energy exchange between the fluids in Sec.~II.B. In Sec.~III.A we present
the linearized dynamical equations, and in  Sec.~III.B we derive the
instability criterion for the condensation mode. The analysis of the
instability is given in Sec.~IV. We close our 
paper with a short summary in Sec.~V.

\section{DYNAMICS AND THERMODYNAMICS}

\subsection{Dynamical equations}

We start from a four-fluid unmagnetized system (based on
Refs.~\onlinecite{ksbs97} and \onlinecite{schroeer98} for a magnetized plasma)
for electrons ($\alpha$=$e$), ions ($\alpha$=$i$), neutral atoms ($\alpha$=$a$)
and massive charged dust particles ($\alpha$=$d$):

\be
\label{conti}
\de_tn_\alpha+\div (n_\alpha {\bf u}_\alpha)=0,
\ee

\be
\label{mom}
\de_t(n_\alpha{\bf u}_\alpha)+\div (n_\alpha{\bf u}_\alpha
{\bf u}_\alpha)=-\frac{1}{m_\alpha}\grad p_\alpha
-{n_\alpha \tilde{Z}_\alpha e\over m_\alpha}\grad\Phi,       
\ee

\be
\label{energy}
\de_t p_\alpha=-\nabla\cdot(p_\alpha{\bf u}_\alpha)+(\gamma-1)
(-p_\alpha\div{\bf u}_\alpha+{\cal H}_\alpha),
\ee

\be
\label{poisson}
\Delta\Phi=-4\pi e\sum \tilde{Z}_\alpha n_\alpha,
\ee
where $n_\alpha$, $\tilde{Z}_\alpha$, ${\bf u}_\alpha$ and $p_\alpha$ are
particle number density, electrical charge number, velocity and gas pressure
of the $\alpha$th species, respectively. As shown in Ref.~\onlinecite{ibanez}
for a wide range of wavelengths relevant to the RC instability, separation of
the dust charge can be neglected, such that the charged components are coupled
through the quasi-neutrality condition  

\be
\label{quasi}
n_e=\sum\limits_{\alpha\neq e}^{}\tilde{Z}_\alpha n_\alpha.  
\ee

We assume the ions to be singly charged, so that $\tilde{Z}_i=+1$ and
$\tilde{Z}_e=-1$. As the dust is usually negatively charged in the systems
under consideration, we write $\tilde{Z}_d=-Z_d$, so that a positive value
of $Z_d$ corresponds to negatively charged dust and vice versa. The
quasi-neutrality condition, thus, reads 

\be
\label{quasinew}
n_i-Z_d n_d-n_e=0.
\ee

However, the validity of this assumption is limited by the absolute value of
dust charge: the characteristic time scale for a radiatively cooling plasma
is of the order of $\sim k_BT/\Lambda n\sim 10^{11}n^{-1}$s in temperature
range $T\sim 10^4$~K. Here, $\Lambda\sim 10^{-23}$ erg cm$^3$ s$^{-1}$ is the
radiative cooling function at $T\sim 10^4$~K for the interstellar environment,
and $n$ is the total number density of all species together. Dust separation
becomes important when this time is shorter than the inverse dust plasma
frequency $\omega_{pd}^{-1}=\sqrt{m_d/4\pi Z_d^2e^2n_d}\sim 10^8n^{-1/2}$~s,
for $n_d\sim 10^{-12}n$ typical for interstellar plasma, which gives
$Z_d\ll 10^{-3}\sqrt{n}$. 

For simplicity we neglect here the collisional source terms in the momentum
equations. In the case when all the four fluids in the unperturbed state are
at rest, ${\bf u}_\alpha=0$, we consider here, this assumption can only result
in an overestimate of the growth rate, but cannot change the instability
criterion. 

\subsection{Energy exchange} 

The source terms in the energy equations, ${\cal H}_\alpha=H_\alpha-L_\alpha$,
describe the generalized energy gain rate due to radiation and collisional
processes. In the interstellar medium (ISM) the RC instability develops mostly 
in the warm neutral medium with $T\sim 10^4$ K, $n\sim0.1$ cm$^{-3}$ and a
fractional ionization of $x\sim 0.1$. Small amount of trace heavy elements
account for $\sim 0.01$ of the gas mass, and a similar amount is confined in
dust grains. The sizes of grains range from $\sim 10$\AA\ to 0.4$\mu$m (cf.
Ref.~\onlinecite{draine03}), the charge of dust grains in the warm neutral
medium is positive and varies from $|Z_d|\sim 0.1$ to $|Z_d|\sim 10^2$, e.g.
Ref.~\onlinecite{yan04}, depending on size; for the sake of simplicity we will
describe in what follows the dust component as an ensemble of particles of
equal masses, $m_d=10^{-14}$g, and equal charges, $Z_d$, assuming the average
abundance of dust particles, $x_d=n_d/n\sim 10^{-12}$. The ion component
comprises mostly of protons and singly ionized trace elements such as CII,
FeII, and SiII (ionized by the interstellar ultraviolet field), the neutrals
contain hydrogen atoms, HI, and neutral oxygen, OI. Helium is neutral in the
warm neutral medium, and its contribution to radiative processes is negligible 
at low temperatures $T<2\times 10^4$ K, therefore it can be included in the
system by multiplying the mass of the neutral particles by factor of 1.4. We
have used the two sets of abundances corresponding to the standard solar
abundances $x_i=n_i/n$ for species $i$: $x_{Fe}=3.2\times10^{-5}$,
$x_{Si}=3.2\times10^{-5}$, $x_{O}=4.4\times10^{-4}$, and
$x_{C}=3.57\times10^{-4}$, and a "depleted" set of abundances with
$x_{Fe}=6.0\times10^{-7}$, $x_{Si}=2.0\times10^{-6}$,
$x_{O}=5.0\times10^{-4}$, and $x_{C}=6.0\times10^{-5}$, describing the mean
composition of the interstellar plasma with some elements frozen on dust
grains as suggested first by Ref.~\onlinecite{field9}. The results for
the two sets are similar; later only calculations for the "depleted" set
are shown.

We assume in this paper that the system gains the energy from photo-electrons
produced by external ultraviolet radiation ionizing dust particles. Non-thermal
photoelectrons emitted by dust share their energy with thermal electrons of
the plasma with the rate $\Gamma n_d$, where $\Gamma$ is determined by the
external UV radiation flux, $G_0$, and optical properties of the dust.
In inelastic collisions with ions and neutrals the electrons lose their energy
to excite internal degrees of freedom, which then decay radiatively. In the
simplest case of an optically thin low-density plasma these energy losses by the
electrons can, thus, be written as the sum of two processes: $L_i^e(T_e)n_in_e$
and $L_{a}^e(T_e)n_an_e$, where we implicitly assumed that the thermal velocity
of the electrons $v_{T,e}\gg v_{T,i},~v_{T,a}$, so that in $L_i^e$ and $L_a^e$
only the dependence on $T_e$ is essential. In addition, the electrons can
share their thermal energy in elastic collisions with the ions and neutrals.
The coefficients are $q^e_i$ and $q^e_a$, respectively. In total it gives 

\ba
\label{elec}
{\cal H}_e&\!=\!&\Gamma n_d-L_i^e(T_e)n_in_e-L_a^e(T_e)n_an_e\nonumber\\
&&\phantom{\Gamma n_d}-q_i^e(T_e)\Bigl(T_e-T_i\Bigr)n_in_e\nonumber\\
&&\phantom{\Gamma n_d}-q_a^e(T_e)\Bigl(T_e-T_a\Bigr)n_an_e,
\ea
where for $\Gamma n_d$ we have taken the value typical for the interstellar
plasma illuminated by the galactic ultraviolet (UV) radiation field according
to Ref.~\onlinecite{wolfire}. The cooling functions $L_i^e$ and $L_a^e$ are
taken from Refs.~\onlinecite{wolfire} and~\onlinecite{penston}, the
coefficients for the collisional energy exchange between the electrons and
ions $q_i^e=q_i^e(T_e)$ and the electrons and neutrals $q_a^e=q_a^e(T_e)$ are
given in Refs.~\onlinecite{draine80} and~\onlinecite{draine83}. 

Energy losses of the ions in inelastic collisions are small in comparison with
energy exchange rates in elastic collisions, and these losses are normally
neglected. In our calculations we included, however, these losses, assuming,
according to Ref.~\onlinecite{dalg67}, that the cross section for direct
excitation by proton impact is broadly similar to the one by electron impact
if the proton has the same velocity as the electron, with the only difference
that the maximum cross section by proton impacts is a factor of 10 larger than
that by electrons (cf. Refs.~\onlinecite{bates53,carew, fite61}). With this
assumption the derived form of the radiative cooling functions for the ions
read as $L_i^i\simeq \alpha (m_e/m_i)^{3/2}L_i^e$ for the ion-ion inelastic
collisions and $L_a^i\simeq \alpha(m_e/m_i)^{3/2}L_a^e$ for the collisions
between the ions and neutrals, where we varied the factor $\alpha$ from 1 to
10. Therefore, we adopt for the ions the generalized rate in the form 

\ba
\label{ion}
{\cal H}_i&=&-L_i^i(T_i){n_i}^2-L_a^i(T_i)n_an_i-q_d^i(T_i)T_in_in_d\nonumber\\
&&-q_i^e(T_e)\Bigl(T_i-T_e\Bigr)n_in_e\nonumber\\
&&-q_a^i(T_a,T_i)\Bigl(T_i-T_a\Bigr)n_in_a,
\ea
where $q_a^i$ depends on both $T_i$ and $T_a$ and is given in
Refs.~\onlinecite{draine80} and~\onlinecite{draine83}; $q_d^i$ describes
cooling of the ions in elastic collisions with translationally cold dust
particles -- by the order of magnitude $q_d^i\sim 2k_B(m_p/m_d)
\pi {\sigma_d}^2v_{T_i}$, where $m_p/m_d$ is the mass ratio of the ions
(mainly protons) to dust particles, $\sigma_d$ the geometric cross-section of
a dust particle, $v_{T_i}$ the thermal velocity of the ions, and is small
compared to other terms in (\ref{ion}); the second term on the r.h.s.
describes radiative losses of the ions in collisions with neutrals; the
contribution of ion-ion inelastic collisions in a relatively weakly ionized
plasma we deal with ($x_e\simlt0.1$) is small.

Similarly we obtain for the neutrals 

\ba
\label{neut}
{\cal H}_a&\!=\!q&-L_i^a(T_a)n_in_a-L_a^a(T_a)n_a^2-q_d^a(T_a)T_an_an_d\nonumber\\
&&-q_a^i(T_a,T_i)\Bigl(T_a-T_i\Bigr)n_an_i\nonumber\\
&&-q_a^e(T_e)\Bigl(T_a-T_e\Bigr)n_an_e,
\ea
where the first term in the r.h.s. describes radiative energy losses of the 
neutral atoms (mostly hydrogen) in collisions with the ions (mostly the ions 
of heavy elements, like CII, FeII), while the second term corresponds to
energy losses in collisions of mostly hydrogen atoms with such heavy elements
as OI. The third term in (\ref{neut}) describes cooling of the neutrals in
elastic collisions with presumably cold dust particles -- loosely $q_d^a\sim
q_d^i$ and both contributions are small in comparison with other terms and are,
thus neglected in the results shown below.

In the equilibrium state of the component the following set of equations
are fulfilled
\be
\label{teq}
{\cal H}_e=0,~{\cal H}_i=0,~{\cal H}_a=0, 
\ee
where the energy gain for the ions and neutrals stems from a small 
difference in temperatures $T_e> T_i,~T_a$, such that $T_e-T_i,~
T_e-T_a\ll T_e$ and approximate equalities $T_e\simeq T_i\simeq T_a$ 
hold in the whole temperature range. The solution of (\ref{teq}) 
is depicted as dotted lines on Figs 1-3. 

\section{STABILITY ANALYSIS}

\subsection{Linear perturbations}

We consider planar perturbations, $\delta g$, of a quantity $g$ in the form
$\delta g\propto\exp[i(kx-\omega t)]$, i.e. we assume the perturbations to
propagate only in $x$-direction for simplicity. In this case the equations
(\ref{conti}) to (\ref{poisson}) read 

\be
\label{contil}
\omega \nu_\alpha- k u_\alpha=0,
\ee

\be
\label{moml}
\omega u_\alpha=kc_\alpha^2\nu_\alpha+kc_\alpha^2\tht_\alpha
+{\tilde{Z}_\alpha e\over m_\alpha}k\Phi,
\ee

\ba
\label{energyl}
-i\omega\nu_\alpha-i\omega\tht_\alpha=-i\gamma ku_\alpha\!\!\!\!\!\!
&&+\sum\limits_{\beta}^{}{1\over T_{\alpha,0}}{\de h_\alpha
\over \de n_\beta}{n_{\beta,0}\over n_{\alpha,0}}\nu_\beta\nonumber\\
&&+\sum_{\beta}^{}{1\over n_{\alpha,0}}{\de h_\alpha\over \de T_\beta}
{T_{\beta,0}\over T_{\alpha,0}}\tht_\beta,
\ea

\be
\label{pois}
\Phi={4\pi e\over k^2}\sum \tilde{Z}_\alpha n_\alpha\nu_\alpha,
\ee
where $h_\alpha=(\gamma-1){\cal H}_\alpha/k_B$, $\nu_\alpha=\delta n_\alpha/
n_{\alpha,0}$, $\tht_\alpha=\delta T_\alpha/T_{\alpha,0}$, $k_B$ is the
Boltzmann constant. Combining (\ref{contil}), (\ref{moml}), and (\ref{energyl})
one obtains

\be
\label{syst1}
i(\gamma-1)\omega\nu_\alpha-\sum_\beta^{}\biggl(i\omega\eta_{\alpha\beta}
+\lambda_{\alpha\beta}+\sum_\kappa^{}\mu_{\alpha\kappa}\eta_{\kappa\beta}
\biggr)\nu_\beta=0,
\ee
where $\alpha$ and $\kappa$ run over the lighter species $i$, $e$, and
$a$, whereas index $\beta$ runs over all four species, i.e. $i$, $e$, $a$, and
$d$. The matrix $\eta_{\alpha\beta}$ is given in Appendix A. This equation
must be complemented by a condition that follows from inserting Eq.
(\ref{pois}) into Eq. (\ref{moml}) for the (cold) dust:

\be
\label{syst2}
Z_d{m_e\over m_d}\omega_{pe}^2\nu_e-Z_d{m_i\over m_d}\omega_{pi}^2\nu_i
+(\omega_{pd}^2-\omega^2)\nu_d=0,
\ee
where $\omega_{p\alpha}$ is the plasma frequency of the $\alpha$th component.
Further we define the abbreviations

\be
\label{notat}
\lambda_{\alpha\beta}={1\over T_{\alpha,0}}{\de h_\alpha\over \de n_\beta}
{n_{\beta,0}\over n_{\alpha,0}},~
\mu_{\alpha\beta}={1\over n_{\alpha,0}}{\de h_\alpha\over \de T_\beta}
{T_{\beta,0}\over T_{\alpha,0}}.
\ee
Explicit expressions for $\lambda_{\alpha\beta}$ and $\mu_{\alpha\beta}$ are
given in Appendix B. We concentrate in this paper on the case when the
external (unperturbed) electric and magnetic fields are zero, and, thus, in
linear approximation the Lorentz force can be neglected. 

\subsection{The instability criterion} 

From system (\ref{syst1}) and Eq. (\ref{syst2}) we obtain the
following dispersion, where we use the abbreviation $N=-i\omega$ for the
growth rate:

\begin{equation}
\label{disp}
\left|
\begin{array}{cccc}
D_iN\!\!+\!\!\tilde\Lambda_{ii}&-\eta_{ie}N\!\!+\!\!\tilde\Lambda_{ie}&
\tilde\Lambda_{ia}&-\eta_{id}N\!\!+\!\!\tilde\Lambda_{id}\\
-\eta_{ei}N\!\!+\!\!\tilde\Lambda_{ei}&D_eN\!\!+\!\!\tilde\Lambda_{ee}&
\tilde\Lambda_{ea}&-\eta_{ed}N\!\!+\!\!\tilde\Lambda_{ed}\\
\tilde\Lambda_{ai}&\tilde\Lambda_{ae}&\!\!\!\!D_aN\!\!+\!\!\tilde\Lambda_{aa}\!\!\!\!&\tilde\Lambda_{ad}\\
-\frac{m_i}{m_d}Z_d\omega_{pi}^2&\frac{m_e}{m_d}Z_d\omega_{pe}^2&0&N^2\!\!+\!\!\omega_{pd}^2 
\end{array}
\right|=0,
\end{equation}
with $\tilde\Lambda_{\alpha\beta}=\lambda_{\alpha\beta}+\sum_\kappa^{}
\mu_{\alpha\kappa}\eta_{\kappa\beta}$, $D_\alpha=(\gamma-1)
-\eta_{\alpha\alpha}$. In the limit $N^2\ll \omega_{pd}^2$ the charge
separation becomes unimportant and (\ref{disp}) converges to 
\begin{equation}
\label{disp1}
\left|
\begin{array}{cccc}
D_iN+\Lambda_{ii}& \Lambda_{ie}& 
\Lambda_{ia}& \lambda_{id}\\
\Lambda_{ee}& D_eN+\Lambda_{ee}& 
\Lambda_{ea}& \lambda_{ed}\\
\Lambda_{ai}& \Lambda_{ae}& 
D_aN+\Lambda_{aa}& \lambda_{ad}\\
-n_i/Z_dn_d& n_e/Z_dn_d&0& 1 
\end{array}
\right|=0, 
\end{equation}
where $\Lambda_{\alpha\beta}=\lambda_{\alpha\beta}+\mu_{\alpha\beta}d_\beta$,
$d_\beta=\omega^2/k^2c_\beta^2-1$. We assumed here explicitly that the heating
and cooling rates of the plasma components $e,~i,~a$ do not depend on the dust
kinetic temperature, $T_d$, which implies $\mu_{ed}=\mu_{id}=\mu_{ad}=0$.
Equation (\ref{disp1}) is the dispersion relation for a radiating plasma, when
quasi-neutrality can be assumed. 

In the low-frequency limit $|N|\ll kc_e,~kc_i,~kc_a$ the coefficients
$D_\alpha$ converge to $D_\alpha=\gamma$, and therefore the sufficient
condition for the instability is that the constant term in the polynomial
(\ref{disp1}) in $N$ is negative. Thus, the instability condition is 

\begin{eqnarray}
\label{instab0}
-{n_e\over Z_dn_d}\Bigl(\!\!\!&\phantom{-}&\!\!\!\lambda_{ad}\Lambda_{ei}'
\Lambda_{ia}'+\lambda_{id}\Lambda_{ea}'\Lambda_{ai}'+\lambda_{ed}
\Lambda_{ii}'\Lambda_{aa}'\nonumber\\
\phantom{-n_e\over Z_dn_d}\!\!\!&-&\!\!\!\lambda_{ed}\Lambda_{ai}'\Lambda_{ia}'
-\lambda_{ad}\Lambda_{ii}'\Lambda_{ea}'
-\lambda_{id}\Lambda_{ei}'\Lambda_{aa}'\Bigr)\nonumber\\
-{n_i\over Z_dn_d}\Bigl(\!\!\!&\phantom{-}&\!\!\!\lambda_{ad}\Lambda_{ee}'\Lambda_{ia}'
+\lambda_{id}\Lambda_{ae}'\Lambda_{ea}'+\lambda_{ed}\Lambda_{ie}'\Lambda_{aa}'\nonumber\\
\phantom{-{n_i\over Z_dn_d}}\!\!\!&-&\!\!\!\lambda_{ed}\Lambda_{ia}'\Lambda_{ae}'
-\lambda_{ad}\Lambda_{ie}'\lambda_{ea}'
-\lambda_{id}\Lambda_{ee}'\Lambda_{aa}'\Bigr)\nonumber\\
+\Bigl(\!\!\!&\phantom{-}&\!\!\!\Lambda_{ee}'\Lambda_{ii}'\Lambda_{aa}'+
\Lambda_{ae}'\Lambda_{ei}'\Lambda_{ia}'+
\Lambda_{ie}'\Lambda_{ea}'\Lambda_{ai}'\nonumber\\
\phantom{+(}\!\!\!&-&\!\!\!\Lambda_{ae}'\Lambda_{ea}'\Lambda_{ii}'-
\Lambda_{ie}'\Lambda_{ei}'\Lambda_{aa}'-
\Lambda_{ai}'\Lambda_{ia}'\Lambda_{ee}'\Bigr)\nonumber\\
<0,\phantom{X}&&
\end{eqnarray}
here $\Lambda_{\alpha\beta}'=\lambda_{\alpha\beta}-\mu_{\alpha\beta}$. This
condition is an extension of the standard (one-fluid) condition for the 
condensation mode to grow as given by Ref.~\onlinecite{field}:

\be
{\de {\cal H}\over \de n}<{T_0\over n_0}{\de {\cal H}\over \de T}.
\ee

\section{RESULTS}  

In a one-fluid description under the conditions considered here: $H\propto
n_d \propto n$ and $L=\Lambda(T)n^2$ with fixed (temperature independent)
ionization states, the instability criterion reads as $d\ln\Lambda/d\ln T<1$
and restricts a density independent temperature range where the plasma is
unstable: for the diffuse warm neutral phase of the ISM the instability
can occur in the temperature range $100\simlt T\simlt 7000$ K. The situation
changes dramatically, when a possible separation in the dynamics of the four
plasma components is taken explicitly into account. For the sake of simplicity 
we consider first constant dust heating rate $\Gamma$=const in the whole 
range of temperature and density. This suggests that the dust charge 
and correspondingly the fraction of the UV radiation transformed into gas
heating are kept constant  in the whole range of plasma parameters. It should
be stressed though  that at realistic conditions in the interstellar plasma
dust charge,  and as a result the heating rate, depend on temperature and
density, cf. Ref.~\onlinecite{tielens}: $Z_d=Z_d(T,n_e)$, $\Gamma=
\Gamma(T,n_e)$. However, we believe it worthwhile to restrict ourselves with a
simplified scheme in order to understand better how dust particles affect the
dynamics of the RC instability in a four-fluid description. In the models
shown below we assume also a constant fractional ionization, $x_e=$const. This
assumption is justified by the fact that the instability growth time is
expected to be of the order of the radiative cooling time, which is always
shorter than the relaxation time for recombination (see
Ref.~\onlinecite{dopita}); for conditions, e.g., in the warm phase of the
interstellar medium with $T\sim 10^4$ K the radiative cooling time is
$\sim 10^{11}n^{-1}$ s$^{-1}$, while the recombination time is
$\sim 10^{12}n^{-1}$ s$^{-1}$.

Figure 1 shows the domains of instability in the temperature-density plane 
for several values of the dust charge. Panel a) corresponds to the standard
one-fluid case for a constant fractional ionization $x_e=n_e/n$=0.1: the whole
plane is separated onto two regions independent of the density, where the
plasma is stable (white) or unstable (gray), corresponding to the criterion
$d\ln\Lambda/d\ln T<1$. Panel b) shows the instability domain in the
four-fluid description with neutral dust particles $Z_d=0$: the domain is
clearly density-independent and slightly wider than on the panel for the
one-fluid case -- this corresponds to the condition $d\ln\Lambda/d\ln T<2$,
which holds either when radiative losses are unbalanced by heating, or when
heating is independent of the density, i.e. $H=$const. As we neglect in this
study collisional friction between the components, the dust component is
decoupled from the plasma when $Z_d=0$, which is equivalent to the latter
case, i.e. $\delta H=0$.

\begin{figure}
\begin{center}
\includegraphics[width=1.00\linewidth]{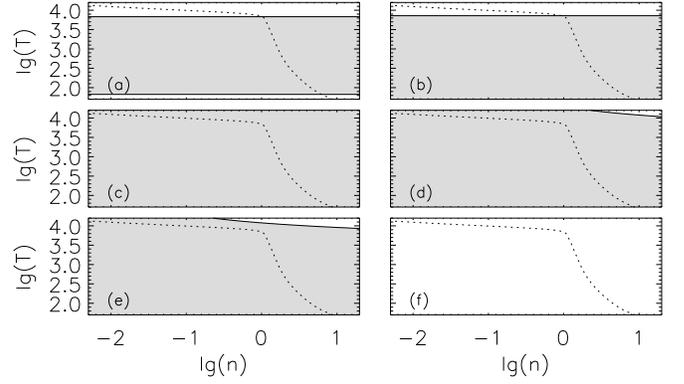}
\end{center}
\caption{The instability criterion in the temperature-density plane
in a double logarithmic representation. Grey areas depict unstable regions,
white corresponds to stable perturbations; the dotted line is the solution
of the equilibrium condition $\Gamma=\Lambda(T)n$ written in the one-fluid
approximation and describes the equilibrium equation of state $T=T(n)$. 
Panel a) shows the separation of the temperature-density plane into
stable and unstable domains for the standard one-fluid case), panel b)
depicts the condition for $Z_d=0$ (b). The middle row shows the shows the
separation for positive dust with $Z_d=-1$ (c) and $Z_d=-10$ (d). The
bottom row shows the cases  $Z_d=-100$ (e) and for negative ($Z_d=0.1$) dust
(f). Note that all cases for negative dust look almost identical in the
plane shown here.
} 
\end{figure}
\noindent 

The panels c), d), and e) correspond to positively charged dust grains,
$Z_d<0$ with increasing absolute values (given in the Figure caption) of
the dust charge. It is clearly seen that positively charged dust particles
destabilize the system, widening the instability domain to higher
temperatures where formally in one-fluid approximation $d\ln\Lambda/d\ln T>1$. 

Negative dust particles, shown in panel f), strongly stabilize the system even 
in the limit of small $Z_d$ -- dust charges in the range $Z_d=+[0.01-100]$ 
give identically stability in the whole domain of the temperature-density
plane shown in Fig. 1. Stabilization of the thermal instability by negatively
charged dust can be understood as follows: due to the quasineutrality
condition negative dust particles repel the electrons out of the compressed
plasma, and thus the radiative cooling rate provided basically by the
electrons decreases. Only when dust charge is high in absolute value,
$Z_d=10^3$, a narrow region at temperatures $100~{\rm K}\simlt T\simlt 10^4$
K, where the instability condition is fulfilled, does appear in 
the high density range as seen in Figure 2. 

\begin{figure}
\begin{center}
\includegraphics[width=0.50\linewidth]{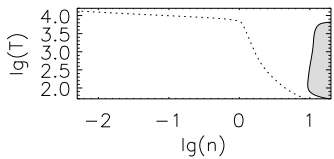}
\end{center}
\caption{ The instability criterion on the temperature-density plane 
for highly charged negative dust grains: $Z_d=10^3$. 
} 
\end{figure}

\begin{figure}
\begin{center}
\includegraphics[width=1.00\linewidth]{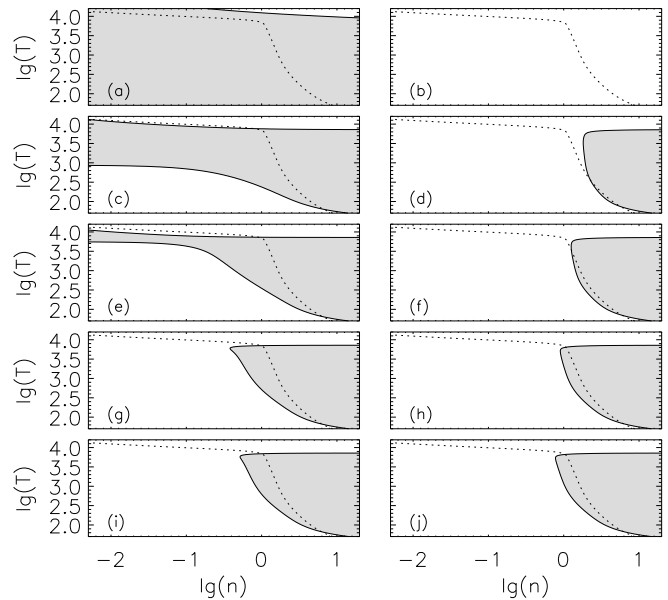}
\end{center}
\caption{ The instability criterion in the temperature-density plane 
for enhanced rate of elastic energy exchange between electrons and ions; 
the left and right panels show the instability domain for positively 
($Z_d=-10^3$) and negatively ($Z_d=10^3$) charged grains, respectively.
The collisional frequency is multiplied by a factor of 0.01 (a,b),
10 (c,d), 20 (e,f), 50 (g,h), 100 (i,j). 
} 
\end{figure}

This instability at high densities in turn seems to stem from a tighter 
collisional coupling between the electrons and ions. Figure 3 shows 
the dependence of the instability domains in the temperature-density 
plane for an artificially enhanced rate of the elastic energy exchange  
between electrons and ions; the left panel corresponds to positively 
charged grains, the right panel to negatively charged grains; in both 
cases the dust charge is fixed at the level $|Z_d|=10^3$. It is readily 
seen that weakening of the elastic energy exchange in collisions between 
the electrons and ions results in a widening of the instability domain 
for positively charged grains (left upper panel), and in a shrinking 
of the instability domain for negatively charged grains (right upper 
panel). More frequent electron-ion collisions lead to a narrower 
instability domain for positive dust, while in the case of negative 
dust the instability domain widens with the collisional rate. In the 
limit of very high collisional frequencies (increased by a factor of
100 with respect to the normal value) the instability domains for positive
and negative grains become almost identical. This means that the instability
domain at high densities for negatively charged grains shown in Figure 2
arises due to elastic interactions between the electrons and ions. The
dependence of the instability domain on the density emerges from the fact
that the energy exchange in elastic collisions between the species does not
appear in the equilibrium energy equation because of the isothermality of
the equilibrium $T_e=T_i=T_a$, while in the instability criterion
(\ref{instab0}) the derivatives of the elastic collisional rates over
temperatures, proportional to the density square, do contribute; at high
densities this contribution becomes dominant. 

For a constant dust charge $Z_d$ variations of the UV radiation flux,
$G_0$, result in a simple shift of the instability domain and the equilibrium
temperature curve along the $n$-axis -- this is a consequence of the fact that
at such an assumption the heating rate linearly depends on density $n$, while
the cooling rate is proportional to the square of density $n^2$: the higher is
$G_0$ the larger is the density needed for radiation losses to balance the
heating at given temperature. This behavior may change to some extent when
the dependence of dust charge on $G_0$ is accounted.

\section{SUMMARY}

The radiation-condensation (thermal) instability in a dusty plasma heated
through photo-ionization of dust particles is described in a four-fluid
approximation. It is shown that in this approximation with admitted separate
motions of the species, the instability criterion changes dramatically
compared to the standard one-fluid case. In particular, positively charged
dust particles are found to destabilize the system, while negatively charged
ones strongly stabilize perturbations: only in the limit $Z_d\gg 1$ negative
dust leaves  a relatively narrow domain of instability in the high density
end, where elastic energy exchange between the electrons and ions dominates. 

In the interstellar medium the dust charge in low-density regions (diffuse
neutral HI phase) is positive due to photo-ionization by stellar ultraviolet
light, cf. Ref.~\onlinecite{yan04}. In the light of our results this means
that low-density intercloud gas is highly unstable against the formation of
condensations in the whole temperature range. A significant fraction of dust
is also positive in denser regions, such as diffuse HI clouds and even in
molecular clouds (see Ref.~\onlinecite{yan04}). Therefore, contrary to a
common understanding that dense phases of the interstellar gas are stable
against the formation of condensations, they turn out to be unstable when
separated motions of the species are accounted. One may speak thus about
possible fragmentation of interstellar clouds through the RC instability on
smaller structures. 

\appendix

\section{$\eta_{\alpha\beta}$}

The matrix $\eta_{\alpha\beta}$ is determined from the momentum 
equations for the components and connects the dimensionless temperature 
and density 

\begin{equation}
\tht_\alpha=\sum_\beta^{}\eta_{\alpha\beta}\nu_\beta
\end{equation}
The elements of the matrix are 

\ba
&&\!\!\!\!\!\!\!\left(
\begin{array}{cccc}
\eta_{ii}&\eta_{ie}&\eta_{ia}&\eta_{id}\\
\eta_{ei}&\eta_{ee}&\eta_{ea}&\eta_{ed}\\
\eta_{ai}&\eta_{ae}&\eta_{aa}&\eta_{ad}\\
\eta_{di}&\eta_{de}&\eta_{da}&\eta_{dd}
\end{array}
\right)=\nonumber\\
&&\!\!\!\!\!\!\!\left(
\begin{array}{cccc}
{\omega^2-\omega_{pi}^2\over k^2c_i^2}-1&{m_e\over m_i}{\omega_{pe}^2\over k^2c_i^2}
&0&{m_d\over Z_dm_i}{\omega_{pd}^2\over k^2c_i^2}\\
{m_i\over m_e}{\omega_{pi}^2\over k^2c_e^2}&{\omega^2-\omega_{pe}^2\over k^2c_e^2}-1
&0&-{m_d\over Z_dm_e}{\omega_{pd}^2\over k^2c_e^2}\\
0&0&\!\!\!{\omega^2\over k^2c_a^2}-1\!\!\!&0\\
0&0&0&0
\end{array}
\right)\!\!.
\ea

\section{$\lambda_{\alpha\beta}$ and $\mu_{\alpha\beta}$}

For simplicity we assume further that $\Gamma$ does not depend on $n_e$ and
$T_e$. In addition, we assume that in equilibrium $|T_e-T_i|,~|T_e-T_a|,
~|T_i-T_a|\ll T_e$. The collision coefficients with the dust, $q^i_d$ and
$q^a_d$ give here, were neglected in the results shown. The coefficients are:

\begin{eqnarray}
\lambda_{ii}&=&-\frac{2L_i^in_i+L_a^in_a}{T_i}+q_d^in_d\nonumber\\
\lambda_{ie}&=&\phantom{-}0\nonumber\\
\lambda_{ia}&=&-\frac{L_a^i}{T_i}n_a\nonumber\\
\lambda_{id}&=&-q_d^in_d
\end{eqnarray}
\begin{eqnarray}
\lambda_{ei}&=&-\frac{L_i^e}{T_e}n_i\nonumber\\
\lambda_{ee}&=&-\frac{L_i^en_i+L_a^en_a}{T_e}\nonumber\\
\lambda_{ea}&=&-\frac{L_a^e}{T_e}n_a\nonumber\\
\lambda_{ed}&=&-\frac{\Gamma n_d}{T_e n_e}
\end{eqnarray}
\begin{eqnarray}
\lambda_{ai}&=&-\frac{L_i^an_i}{T_a}\nonumber\\
\lambda_{ae}&=&\phantom{-}0\nonumber\\
\lambda_{aa}&=&-\frac{L_i^an_i+2L_a^an_a}{T_a}-q_d^an_d\nonumber\\
\lambda_{ad}&=&-q_d^an_d
\end{eqnarray}
\begin{eqnarray}
\mu_{ii}&=&-\frac{dL_i^i}{dT_i}n_i
-\frac{dL_a^i}{dT_i}n_a-\frac{dq_d^i}{dT_i}n_d\nonumber\\
&&-q_i^en_e-q_a^in_a-q_d^i n_d,\nonumber\\
\mu_{ie}&=&\phantom{-}q_i^e n_e,\nonumber\\
\mu_{ia}&=&\phantom{-}q_a^i n_a,\nonumber\\
\mu_{id}&=&\phantom{-}0
\end{eqnarray}
\begin{eqnarray}
\mu_{ei}&=&q_i^en_i,\nonumber\\
\mu_{ee}&=&-\frac{dL_i^e}{dT_e}n_i-\frac{dL_a^e}{dT_e}n_a\nonumber\\
&&-q_i^en_i-q_a^en_a,\nonumber\\
\mu_{ea}&=&\phantom{-}q_a^e n_a,\nonumber\\
\mu_{ed}&=&\phantom{-}0
\end{eqnarray}
\begin{eqnarray}
\mu_{ai}&=&\phantom{-}q_a^in_i,\nonumber\\
\mu_{ae}&=&\phantom{-}q_a^en_e,\nonumber\\
\mu_{aa}&=&-\frac{dL_i^a}{dT_a}n_i-\frac{dL_a^a}{dT_a}n_a
-\frac{dq_d^a}{dT_a}n_d\nonumber\\
&&-q_a^in_i-q_a^en_e-q_d^an_d,\nonumber\\
\mu_{ad}&=&0.
\end{eqnarray}

We accounted here that direct injection of energy from dust 
particles to the ions and neutrals is zero.

\section*{Acknowledgements}

We are grateful for valuable discussions with R.-J. Dettmar and M. Tokar and
thank A. Mikhajlov for his consultation on atomic processes. This work was
supported by the German Science Foundation (DFG) through the
Sonderforschungsbereich 591; YS is partly supported by RFBR (project codes
05-02-17070 and 06-02-16819) and by the Federal Agency of Education of
Russian Federation (project code RNP 2.1.1.3483).

\end{document}